\documentclass[12pt]{article}

\usepackage{amsfonts,amssymb}

\input epsf

\title{Results of numerical simulations \\ for unstable-particles
pair production \\ in modified perturbation theory in
NNLO\thanks{Contribution to Proceedings of {\it XXth International
Workshop HEPQFT'11}, September 24 - October 1,  2011
Sochi, Russia}}

\author{M. L. Nekrasov \\
\small{\it Institute for High Energy Physics, 142281 Protvino, Russia}}

\date{}
\begin{document}

\maketitle

\begin{abstract}
We consider pair production and decay of fundamental unstable
particles in the framework of a modified perturbation theory (MPT),
which treats resonant contributions of unstable particles in the sense
of distributions. The cross-sections for the top-quark pair production
and for the $W$-boson pair production in $e^{+}e^{-}$ annihilation are
calculated within the NNLO in models that admit exact solutions.
In both cases an excellent convergence of the MPT is detected at the
energies close to and above the maximum of the cross section. In the
case of $W$-boson pair production a precision of the description at the
ILC energies is ensured at the level of one per-mille or higher.
\end{abstract}

The processes of pair production and decays of unstable fundamental
particles, such as top quarks and $W$ bosons, play an important role for
testing the Standard Model and for searching physics beyond. In the case
of colliders subsequent to LHC a description of such processes must be
made generally with the NNLO accuracy. This implies that not only the
gauge cancellations and unitarity should be maintained, but also
suitably high accuracy of computation of resonant contributions must be
provided. Unfortunately, the existing methods can provide only the NLO
precision of the description of the cross-section. This is the case with
the double pole approximation (DPA) successfully applied at LEP2
\cite{DPA} or with the complex-mass scheme (CMS) \cite{CMS} intended
mainly to ILC \cite{ILC}. The pinch-technique method, another for a long
time developed approach, in principle can provide the NNLO precision,
but to maintain the gauge cancellations it requires a huge volume of
calculations of extra contributions that pertain formally to the next
level of the precision, which is impractical \cite{Ditt}. So alternative
approaches are required for systematic calculations at the NNLO. A
modified perturbation theory (MPT) \cite{F1,N1,N2} is a suitable
approach for solving this problem. Its main feature is the direct
expansion of the probability instead of amplitude in powers of the
coupling constant with the aid of distribution-theory methods. As the
expansion is made in powers of the coupling constant and the
object to be expanded is gauge invariant, the gauge cancellations
in the MPT must be automatically maintained. Nevertheless, the
accuracy of the description in MPT requires an examination. In order to
do that, numerical simulations are necessary in the framework of MPT.

Since in the case of pair production of unstable particles the most
crucial are the double-resonant contributions, we consider initially
only these contributions in the framework of the model simulations.
(Generally, the single-resonant contributions may be considered, as
well, this is not a problem in the MPT approach \cite{N2}.) In the case
of $e^+ e^-$ annihilation, the corresponding total cross-section has the
form of a convolution of hard-scattering cross-section with the flux
function. The hard-scattering cross-section is an integral over the
virtualities of unstable particles of exclusive cross-section multiplied
by factor
standing for soft massless-particles contributions,
\begin{equation}\label{not1}
\sigma (s) = \int_{s_{\mbox{\tiny min}}}^s \frac{\mbox{d} s'}{s} \:
\phi(s'/s;s) \> \hat\sigma(s')\,, \quad \hat\sigma (s') = \int
\!\!\!\! \int \mbox{d} s_1 \, \mbox{d} s_2 \;
 \hat\sigma_{excl} (s',s_1,s_2) \left(1\!+\!\delta_{soft}\right) \,.
\end{equation}
The exclusive cross-section $\hat\sigma_{excl}$ is written as a product
of Breit-Wigner factors $\rho(s_i)$, some kinematic factors,
and a function $\Phi$, which is the rest of the amplitude squared,
\begin{equation}\label{not2}
\hat\sigma_{excl} (s,s_1,s_2) =
 \theta(\!\sqrt{s}-\!\sqrt{s_1}-\!\sqrt{s_2}\,)
 \sqrt{\lambda (s,s_{1},s_{2})}\;\Phi(s;s_1,s_2) \,
 \rho(s_{1}) \> \rho(s_{2})\,.
\end{equation}
Generally $\Phi$ corresponds to one-particle irreducible
contributions, and so it does not have singularities on the
mass-shell of unstable particles. On the contrary, the kinematic
factors, which include the theta-function and the square root of the
kinematic function $\lambda$, have singularities. The BW factors if
to naively expand them in powers of the coupling constant $\alpha$
generate non-integrable singularities, and this makes up a great
problem because integrals in (\ref{not1}) become senseless.

However, the singularities become integrable if to expand the
B$\!$W factors in the sense of distributions. In this case the
expansion of a separately taken BW factor is beginning with the
$\delta$-function which corresponds to the narrow-width
approximation. The contributions of the naive Taylor expansion are
supplied with the principal-value prescription for the poles. The
nontrivial contributions are the delta-function and its
derivatives with coefficients $c_{n}$, which are polynomials in
$\alpha$ that are determined by the self-energy of the unstable
particle \cite{F1}. Within the NNLO, the expansion is as follows:
\begin{eqnarray}\label{not3}
&\displaystyle \rho(s) \;\;\equiv\;\; \frac{M\Gamma_0}{\pi} \; {|s
- M^2 + \Sigma(s)|^{-2}} \;\;= \;\;\delta(s\!-\!M^2)&
\\
&\displaystyle +\;
    \frac{M \Gamma_{0}}{\pi} \, PV \! \left[\,\frac{1}{(s-M^2)^2}
  -\, \frac{2\alpha\,\mbox{Re}\Sigma_1(s)}{(s\!-\!M^2)^3}\,\right] \;+\;
  \sum\limits_{n\,=\,0}^2 c_{n}(\alpha)\,
  \frac{\mbox{\small ($-$)}^{n}}{n!}\,\delta^{(n)}(s\!-\!M^2) +
  O(\alpha^3)\,.&\nonumber
\end{eqnarray}
Here $M$ is the renormalized mass, $\Gamma_{0}$ is the Born width,
$\Sigma(s)$ is the self-energy of the unstable particle.
Coefficients $c_n$ within the NNLO include 3-loop self-energy
contributions and their derivatives determined on-shell. The
structure of the contributions is such that in the OMS-type
schemes of the UV renormalization the real self-energy
contributions enter into the coefficients either without the
derivatives or with the first derivative only. This means that
the relevant real self-energy contributions are determined by the
renormalization conditions. In the case of unstable particles it
is reasonable to use the $\overline{\mbox{OMS}}$ or pole scheme
\cite{OMS-bar,Sirlin} of the UV renormalization, whose inherent property
is that the renormalized mass of unstable particle by definition
coincides with the real part of the pole of unstable-particle
propagator, which is gauge invariant and scheme-independent. The
coefficients $c_n$ in this scheme are
determined as follows \cite{N2}:
\begin{equation}\label{not4}
c_0  =  - \, \alpha \, \frac{I_2}{I_1} + \alpha^2
\left[\frac{I_2^2}{I^2_1} - \frac{I_3}{I_1} - (I_{1}^{\,\prime})^2
\right],\qquad c_1 = 0, \qquad c_2 = -\, \alpha^2 I^2_1 \,.
\end{equation}
Here $I_{k} = \mbox{Im}\,\Sigma_{k}(M^2)$, $\Sigma = \alpha \,\Sigma_1 +
\alpha^2 \,\Sigma_2 + \alpha^3 \,\Sigma_3$, and $I_{1}^{\,\prime}
= \mbox{Im}\,\Sigma^{\,\prime}_{1}(M^2)$. Simultaneously in the
$\overline{\mbox{OMS}}$ scheme the $\mbox{Im}\,\Sigma(M^2)$ coincides
with the imaginary part of the pole of the propagator. This allows
one to determine $I_{k}$ order-by-order over the width of the
unstable particle via the unitarity relations $\alpha I_1 =
M\Gamma_0$, \ $\alpha^2 I_2 = M \alpha \Gamma_1$, \ and \
$\alpha^3 I_3 = M \alpha^2 \Gamma_2 + \Gamma_0^3/(8M)$~\cite{OMS-bar}.

Unfortunately, expansion (\ref{not3}) has sense only if the weight
in the integral is a regular enough function. In our case,
however, the kinematic factors are not regular, which leads to a
divergence in integrals (\ref{not1}) after the substitution
of the expansions. At first glance, this brings up a question
about the applicability of expansion (\ref{not3}). Nevertheless, the
kinematic factors may be analytically regularized via the substitution
$[\lambda (s,s_{1},s_{2})]^{1/2} \to [\lambda (s,s_{1},s_{2})]^{\nu}$.
With large enough $\nu$ this imparts enough smoothness to the weight,
and the singular integrals become integrable. Fortunately, after
the analytic calculation of integrals the regularization may be
removed without the loss of finiteness of outcomes. Moreover, the
expansion remains asymptotic \cite{N2}. This completely salvages the
applicability of the approach.

The scheme of the analytical calculations is as follows.
At first we proceed to dimensionless energy variables $x$, $x_i$
({\small {\it i} = 1,2}) counted off from thresholds, $\sqrt{s} =
2 M (1 + x/4)$, $\sqrt{s_{i}} = M(1 + x_{i}/2)$. The
hard-scattering cross-section then takes the form
\begin{equation}\label{not5}
 \widetilde{\hat\sigma}(x) = \int\!\!\!\!\int\!\mbox{d} x_1 \, \mbox{d}
x_2 \;
 (x\!-\!x_1\!-\!x_2)_{+}^{\nu} \;
 \widetilde{\Phi}(x\,;x_1,x_2) \;
 \widetilde{\rho}(x_1) \widetilde{\rho}(x_2)\,.
\end{equation}
Here $(x\!-\!x_1\!-\!x_2)_{+}^{\nu} = \theta(x\!-\!x_1\!-\!x_2)
(x\!-\!x_1\!-\!x_2)^{\nu}$ and tilde marks the dimensionless
functions (factor $1\!+\!\delta_{soft}$ is included into the
definition of $\widetilde{\Phi}$). Further, we substitute
asymptotic expansions for $\widetilde{\rho}(x_i)$, and consider at each
$n_i$ ($i=1,2$) the contributions of $PV x_i^{-n_i}$ and
$\delta^{(n_{i}-1)}(x_i)$. Simultaneously, in each case, we represent
the test function in the form of a double Taylor expansion over
$x_i$ truncated at the contributions of $x_i^{(n_i-1)}$, with a
remainder,
\begin{equation}\label{not6}
{\widetilde{\Phi}(x\,;x_1,x_2)} \; = \sum_{k_1=0}^{n_1-1}
\sum_{k_2=0}^{n_2-1}  \frac{x_1^{k_1}}{k_1 !}\;
\frac{x_2^{k_2}}{k_2 !} \:\;
\widetilde{\Phi}^{(k_1,\,k_2)}(x\,;0,0) \;+\;
\Delta\widetilde{\Phi}(x\,;x_1,x_2)\,.
\end{equation}
The higher powers of $x_i$ in the Taylor expansion will zero
the $\delta^{(n_{i}-1)}(x_i)$ and cancel the $PV x_i^{-n_i}$. The
remainder $\Delta\widetilde{\Phi}$ is determined as the difference
between $\widetilde{\Phi}$ and the Taylor expansion. In fact
$\Delta\widetilde{\Phi}$ is to be further expanded with respect to
separately $x_1$ and $x_2$, but for brevity we do not consider here this
procedure explicitly (see details in \cite{N2}). Let us mention
only that the final remainder produces a regular contribution to the
integrand in formula (\ref{not5}), and the integrals of it can be
numerically calculated at $\nu = 1/2$. At the same time, the
contributions of Taylor are singular. However, the integral
(\ref{not5}) of them may be analytically calculated owing to simple
(power) dependence on $x_i$. After making the calculation and after
putting $\nu \to 1/2$, the result appears in the form of a sum of
regular and singular contributions with singular contributions being
products of regular factors and power distributions of the type
$x^{5/2\,-\,n}_{+}$ with integer~$n$. It should be emphasized that at
this stage of calculations the test function $\widetilde{\Phi}$ is
determined by means of conventional perturbation theory, but the
analytic calculations are
made independently of the particular form of $\widetilde{\Phi}$. The
convolution integral of the result can be numerically calculated. In
particular, the integral of singular power distributions can be
calculated by means of the formula
\begin{equation}\label{not7}
\int \mbox{d} x \;\; x_{+}^{\nu} \,\phi(x) =
 \int\limits_{0}^{\infty} \mbox{d} x \;\; x^{\nu}
 \left\{\phi(x) -
 \sum_{k=0}^{N-1} \frac{x^{k}}{k!} \, \phi^{(k)}(0)\right\}\,,
\vspace*{-0.2\baselineskip}
\end{equation}
where $\phi$ is a weight and $N$ is a positive integer such that
$-N\!-\!1< \mbox{Re} \, \nu < -N$.

For carrying out further numerical calculations a double-precision
FORTRAN code is written. The computation of regular integrals in
this code is realized by Simpson method. Numerous indeterminate
forms of the type $0/0$ that emerge in the integrand due to the
difference structures are resolved through the introduction of linear
patches. The patches diminish the errors (numerical instabilities)
that arise because of the loss of decimals near the indeterminacy
points. At the same time, the errors generated by patches may be
numerically estimated. Ultimately, the total errors caused by
indeterminacies may be estimated, as well. The crucial point is
that the errors because of the patches are increasing with increasing
the sizes of the patches, while the errors because of the loss of
decimals are decreasing. So there should be an optimum size of the
patches when the sum of the errors is minimized. The minimization
point must possess extremum properties, so that the result of the
computation at this point must be stable with respect to varying
the sizes of the patches. Furthermore, at the extremum point the
sums of the errors of different kinds must be approximately equal
each other (up to a coefficient of order one). So the order of the
total error may be estimated by the order of the sum of the errors
because of the patches. More details of how to do estimation of
the errors, is found in \cite{N5}. Eventually the adjusted
estimate of the relative error of the computation of the NNLO
approximation turns out to be less than $10^{-3}$ or $10^{-4}$ in two
cases considered below.

The physical models underlying the calculations are related to
processes $e^{+} e^{-} \to \gamma,Z \to t\bar t \to
W^{+}\,b\:W^{-}\,\bar b$ and $e^{+} e^{-} \to \gamma,Z \to
W^{+}\:W^{-} \to 4 f$. Recall that our aim is to verify whether the MPT
calculations are realizable and then to test the convergence properties
of the MPT expansion. Having that in mind, we consider the test function
$\widetilde{\Phi}$ in both cases in the Born approximation. However, the
self-energies in the denominators of propagators of unstable particles,
we consider in 3-loop approximation (see details in \cite{N3,N4}). This
means that we can immediately check by our calculations the convergence
properties of the MPT expansion of the products of BW factors. Actually,
this is sufficient for our purposes because the insertion of the loop
corrections to the test function may be considered as the replacement of
the test function by another one with additional factors $\alpha$,
$\alpha^2$, etc. Meanwhile, as the existing experience shows, the
convergence properties of the MPT expansion are very weakly depend on
the test function, but depend on the values of the corrections to the
widths of unstable particles \cite{N3,N4}. As concerns the soft
massless-particles contributions, we consider among them only universal
ones. They are collected in the flux function and in the Coulomb
factor. The flux function, we take into consideration in the
leading-log approximation. The Coulomb factor, we consider in the
one-gluon/photon approximation with specific resummation
\cite{Coulomb} that does not affect the BW factors. Notice that
although the multi-gluon contributions generally are important in
the case of the top quarks, we believe that at distance from the
threshold a qualitative picture may be simulated in the one-gluon
approximation.

\begin{figure}
\hbox{ 
       \epsfxsize=\textwidth \epsfbox{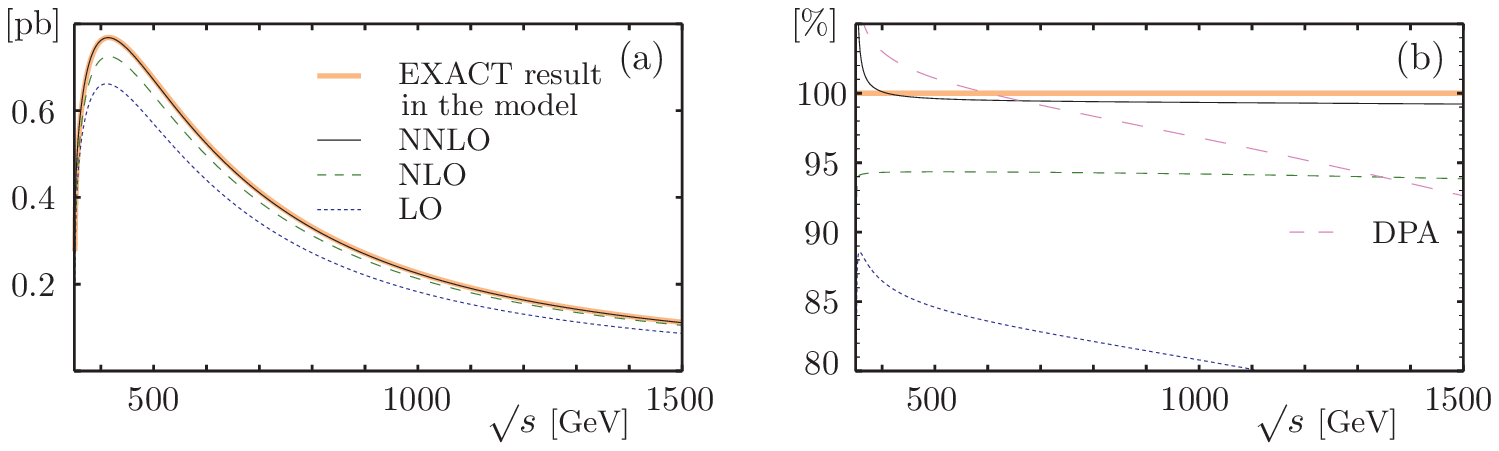}}
\caption{The total cross-section in the model for $t \bar t$ production
and decay.} \label{fig1}
\end{figure}

\begin{figure}
\hbox{ 
       \epsfxsize=\textwidth \epsfbox{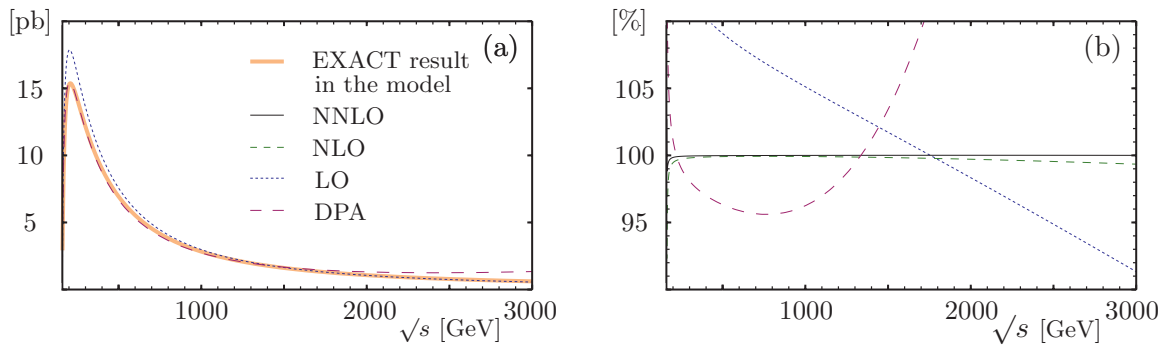}}
\caption{The total cross-section in the model for $W^{+} W^{-}$
production and decay.} \label{fig2}
\end{figure}

\begin{figure}[p]
\begin{center}
\hbox{ \hspace*{55pt}
       \epsfxsize=0.7\textwidth \epsfbox{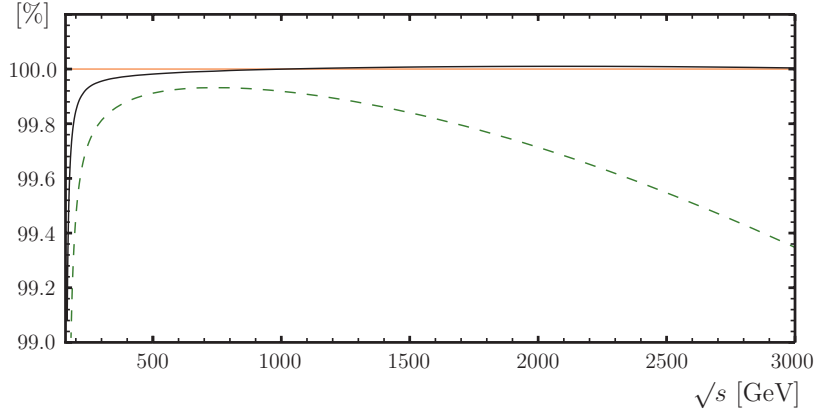}}
\caption{The results of Fig.~2.b with greater scale on vertical
axis.} \label{fig3}
\end{center}
\vspace*{-2.5\baselineskip}
\end{figure}
\begin{table}[p]
\begin{center}
\caption{\small $t \bar t$ production: the total cross-section in
pb and in \% with respect to exact result in the model.} \label{T1}
\begin{tabular}{ c  c c c c  }
\hline\noalign{\medskip} \\[-6mm]
 $\quad \sqrt{s}$ (GeV) $\qquad$
 & $\qquad \sigma_{\,\mbox{\tiny EXACT}} \qquad$      &
 $\qquad \sigma_{LO} \qquad$        & $\qquad \sigma_{NLO} \qquad$ &
 $\quad \sigma_{NNLO}   \qquad$ \\
\hline\noalign{\medskip} \\[-6mm]
 500                               & 0.6724         &
 0.5687          &  0.6344         & 0.6698             \\
                                   & {\small 100\%} &
 {\small 84.6\%} & {\small 94.3\%} & {\small 99.6\%}    \\
\hline\noalign{\medskip} \\[-6mm]
 1000                              & 0.2255         &
 0.1821          &  0.2124         & 0.2240             \\
                                   & {\small 100\%} &
 {\small 80.8\%} & {\small 94.2\%} & {\small 99.3\%}    \\
\hline\noalign{\medskip} \\[-6mm]
 1500                              & 0.1122         &
 0.0867          &  0.1053         & 0.1113             \\
                                   & {\small 100\%} &
 {\small 77.3\%} & {\small 93.8\%} & {\small 99.2\%}    \\[-1mm]
\noalign{\smallskip}\hline
\end{tabular}
\end{center}
\vspace*{-2\baselineskip}
\end{table}

\begin{table}[p]
\begin{center}
\caption{\small $W^{+} W^{-}$ production: the total cross-section
in pb and in \% with respect to exact result in the model.} \label{T2}
\begin{tabular}{ c  c c c c  }
\hline\noalign{\medskip} \\[-6mm]
 $\quad \sqrt{s}$ (GeV) $\qquad$
 & $\qquad \sigma_{\,\mbox{\tiny EXACT}} \qquad$      &
 $\qquad \sigma_{LO} \qquad$        & $\qquad \sigma_{NLO} \qquad$ &
 $\quad \sigma_{NNLO}   \qquad$ \\
\hline\noalign{\medskip} \\[-6mm]
 200                               & 15.258         &
 17.839          &  15.175         & 15.235             \\
                                   & {\small 100\%} &
{\small 116.92\%} &{\small 99.46\%}&{\small 99.85\%}    \\
\hline\noalign{\medskip} \\[-6mm]
 500                               & 6.9355         &
 7.5657          &  6.9294         & 6.9342             \\
                                   & {\small 100\%} &
{\small 109.09\%}&{\small 99.91\%} &{\small 99.98\%}    \\
\hline\noalign{\medskip} \\[-6mm]
 1000                              & 2.8286         &
 2.9733          &  2.8263         & 2.8285             \\
                                   & {\small 100\%} &
{\small 105.12\%}&{\small 99.92\%} & {\small 100.00\%}  \\
\hline\noalign{\medskip} \\[-6mm]
 3000                              & 0.61023        &
 0.55733         &  0.60625        & 0.61026            \\
                                   & {\small 100\%} &
{\small 91.33\%} &{\small 99.35\%} &{\small 100.00\%}   \\[-1mm]
\noalign{\smallskip}\hline
\end{tabular}
\end{center}
\vspace*{-\baselineskip}
\end{table}

The outcomes of computations are presented in the Figures and in
the Tables. In Fig.~\ref{fig1} in the panels (a) the thick curve
shows the behavior of the total cross-section in the model in the
case of the top-quark pair production. The dotted, dashed, and
continuous thin curves show the results of the MPT computations in
the LO, NLO, and NNLO approximations, respectively. It is worth
noting that the NNLO result almost coincides with the exact result
in the region near and above the maximum of the cross-section. The
distinction is visible in the panel (b) where the percentages with
respect to the exact result are presented. In Fig.~\ref{fig2} the
appropriate results are presented in the case of $W$ pair
production. The main difference at Fig.~\ref{fig2} is that 
already the NLO curve almost coincides with the exact result with the
distinction visible in the panel (b) only. In Fig.~\ref{fig2} the
appropriate results are presented in the case of $W$ pair
production. The main difference at Fig.~\ref{fig2} is that already
the NLO curve almost coincides with the exact result with the
distinction visible in the panel (b) only. In Fig.~\ref{fig3} the
results for NLO and NNLO of Fig.~\ref{fig2}.b are repeated
with greater scale on vertical axis. In Tables~\ref{T1} and \ref{T2} the
outcomes are represented in the numerical form at the characteristic
energies at ILC \cite{ILC}. Note that the relative errors of the
computation are less than $10^{-3}$ and $10^{-4}$ in the cases of the
top quarks and $W$ bosons, respectively. Therefore the errors are
omitted in the data of Tables.

In conclusion, first of all we emphasize that the above results
show in practice the existence of the MPT expansion in the case of
pair production and decay of fundamental unstable particles.
Secondly, the NLO and NNLO approximations in the MPT have very stable
behavior at the energies near and above the maximum of the
cross-section. The latter result to a large extent is model-independent
since it is established in different models. (The latter point is
discussed in more details in \cite{N3,N4}. Notice also that at lower
energies, in particular near threshold, the mode of MPT that has been
considered here becomes inapplicable, but there is another mode for MPT
\cite{N2}.) Thirdly, at the ILC energies the NNLO approximation in the
MPT give highly satisfactory results in numerical sense. Namely, in
the case of the top-quark pair production it gives approximately a
half-percent accuracy of the description of the cross-section, and in
the case of $W$ bosons does a per-mille accuracy. In fact, this is what
is needed at the ILC. So, we conclude that MPT is a good candidate for
support at the ILC the pair production and decay of fundamental unstable
particles.

\end{document}